%% file: mmn-mia.tex
\documentclass[times,twocolumn,final]{elsarticle}

\usepackage{medima}
\usepackage{framed,multirow}
\usepackage{amsmath,amssymb,amsfonts}
\usepackage{latexsym}
\usepackage{url}
\usepackage{hyperref}
\usepackage{algorithm}
\usepackage{algpseudocode}
\usepackage{booktabs}

\journal{Medical Image Analysis}
\begin{document}
\verso{Given-name Surname \textit{et~al.}}
\begin{frontmatter}
\title{Self-Supervised Masked Mesh Learning for Unsupervised Anomaly Detection on 3D Cortical Surfaces}

\author{\large{
    Hao-Chun Yang\textsuperscript{1,2,*},
    Sicheng Dai\textsuperscript{1},
    Saige Rutherford\textsuperscript{1,2},
    Christian Gaser\textsuperscript{3},
    Andre F Marquand\textsuperscript{2},\\
    Christian F Beckmann\textsuperscript{2},
    Thomas Wolfers\textsuperscript{1,2,*}}
}

\address{
    \textsuperscript{1}Department of Psychiatry and Psychotherapy, Tubingen Center for Mental Health, Germany\\
    \textsuperscript{2}Donders Institute, Radboud University, Nijmegen, Netherlands\\
    \textsuperscript{3}Department of Psychiatry, University of Jena, Jena, Germany
}

\cortext[cor1]{
    Correspondence to: Calwerstraße 14, 72076 Tübingen, Germany.\\
    \indent\indent E-mail addresses: \href{mailto:chadyang.hc@gmail.com}{chadyang.hc@gmail.com} (H.-C. Yang),\\
    \indent\indent \href{mailto:dr.thomas.wolfers@gmail.com}{dr.thomas.wolfers@gmail.com} (T. Wolfers).}

\received{XX XXX XXXX}
\finalform{XX XXX XXXX}
\accepted{XX XXX XXXX}
\availableonline{XX XXX XXXX}
\communicated{XXX}

\input{0-abstract.tex}
\end{frontmatter}


\input{1-introduction.tex}
\input{2-related.tex}
\input{3-methods.tex}
\input{4-experiments.tex}

\input{5-discussion.tex}
\input{7-conclusion.tex}

\section*{Acknowledgment}\label{sec:acknowledgments}
We would like to thank all participants in this study. This work was supported the German Research Foundation (DFG) Emmy Noether with reference 513851350 (TW) and the BMBF-funded de.NBI Cloud within the German Network for Bioinformatics Infrastructure (031A532B, 031A533A, 031A533B, 031A534A, 031A535A, 031A537A, 031A537B, 031A537C, 031A537D, 031A538A). UKB data were accessed under application number 23668. Data collection and sharing for the Alzheimer's Disease Neuroimaging Initiative (ADNI) is funded by the National Institute on Aging (National Institutes of Health Grant U19AG024904). The grantee organization is the Northern California Institute for Research and Education. In the past, ADNI has also received funding from the National Institute of Biomedical Imaging and Bioengineering, the Canadian Institutes of Health Research, and private sector contributions through the Foundation for the National Institutes of Health (FNIH) including generous contributions from the following: AbbVie, Alzheimer's Association; Alzheimer's Drug Discovery Foundation; Araclon Biotech; BioClinica, Inc.; Biogen; Bristol-Myers Squibb Company; CereSpir, Inc.; Cogstate; Eisai Inc.; Elan Pharmaceuticals, Inc.; Eli Lilly and Company; EuroImmun; F. Hoffmann-La Roche Ltd and its affiliated company Genentech, Inc.; Fujirebio; GE Healthcare; IXICO Ltd.; Janssen Alzheimer Immunotherapy Research \& Development, LLC.; Johnson \& Johnson Pharmaceutical Research \& Development LLC.; Lumosity; Lundbeck; Merck \& Co., Inc.; Meso Scale Diagnostics, LLC.; NeuroRx Research; Neurotrack Technologies; Novartis Pharmaceuticals Corporation; Pfizer Inc.; Piramal Imaging; Servier; Takeda Pharmaceutical Company; and Transition Therapeutics.
\bibliographystyle{model2-names.bst}
\biboptions{authoryear}
\bibliography{1-introduction, 2-related, 3-methods, 4-experiments, 5-discussion, 6-limitations, 7-conclusion}
\end{document}

%% file: 0-abstract.tex
\begin{abstract}
    Unsupervised anomaly detection in brain imaging is challenging. In this paper, we propose self-supervised masked mesh learning for unsupervised anomaly detection on 3D cortical surfaces. Our framework leverages the intrinsic geometry of the cortical surface to learn a self-supervised representation that captures the underlying structure of the brain. We introduce a masked mesh convolutional neural network (MMN) that learns to predict masked regions of the cortical surface. By training the MMN on a large dataset of healthy subjects, we learn a representation that captures the normal variation in the cortical surface. We then use this representation to detect anomalies in unseen individuals by calculating anomaly scores based on the reconstruction error of the MMN. We evaluated our framework by training on population-scale dataset \textit{UKB} and \textit{HCP-Aging} and testing on two datasets of Alzheimer's disease patients \textit{ADNI} and \textit{OASIS3}. Our results show that our framework can detect anomalies in cortical thickness, cortical volume, and cortical sulcus characteristics, which are known to be biomarkers of Alzheimer's disease. Our proposed framework provides a promising approach for unsupervised anomaly detection based on normative variation of cortical features.
\end{abstract}

\begin{keyword}
    \MSC 68T05\sep 68T10\sep 92C55\sep 68U05
    \KWD Unsupervised Anomaly Detection\sep Self-Supervised Learning\sep Brain Imaging\sep Cortical Surface\sep Mesh Convolution
\end{keyword}
    

%% file: 1-introduction.tex
\section{Introduction}\label{sec:introduction}
The detection of dementia through brain imaging has garnered significant attention in recent years \cite{ahmed2018neuroimaging,arvanitakis2019diagnosis}. Surface modeling of the cortex is crucial to advance our understanding of neurological conditions such as dementia \cite{rosen2002patterns,brown2014brain}. The cerebral cortex, with its intricate folds and grooves, offers a wealth of information that goes beyond what traditional volumetric neuroimaging can provide \cite{thomas2015post,grasby2020genetic}. By representing the 3D structure of the cortical surface, researchers can capture detailed morphological features such as cortical thickness, gyrification patterns, and sulcal depth\cite{blanc2015cortical,vuksanovic2019cortical,thompson2007tracking}. These features are crucial in identifying subtle anomalies that may signify the early stages of dementia and cognitive decline \cite{im2008sulcal,tang2021slower}. Techniques such as cortical parcellation, surface reconstruction, and mesh generation allow for precise modeling and analysis of the cortical surface, enabling the detection of fine structural changes that are often missed by broader volumetric approaches. This cortex-based analysis is essential to develop more reliable biomarkers and possibly improve early diagnosis and intervention strategies. The advantages of cortical surface modeling lie in its ability to provide a rich set of geometric features \cite{darayi2022computational,ma2022cortexode}.

Most existing methods that model cortical alteration in disease \cite{rallabandi2020automatic,shin2021cortical,pateria2024comprehensive}, rely on supervised learning approaches applied to neuroimaging data. However, these methods have limitations. Firstly, supervised learning requires large patient and control datasets, which can be challenging and costly to obtain and impossible to obtain for rare diseases \cite{decherchi2021opportunities,dos2022towards,chen2023algorithmic}. Furthermore, such approaches do not fully account for the heterogeneous manifestations of symptoms in different types of complex disorders and diseases characterized by cognitive decline \cite{knopman2003essentials,pike2022subjective}. Secondly, most machine learning and deep learning methods focus on volumetric data \cite{brand2019joint,zhang2022classification,yamanakkanavar2020mri,han2023multi}, with limited exploration of cortical surfaces. The cortical surface exhibits intricate morphological patterns that could provide insight into brain structure and function. Therefore, leveraging an unsupervised approach to model high-dimensional cortical surface data could facilitate the detection of brain anomalies, aiding in the identification of novel disease subtypes and enhancing generalization to new data and populations without relying on labeled supervision.

In this study, we propose an analytical framework, Self-Supervised Masked Mesh Learning for Unsupervised Anomaly Detection on 3D Cortical Surfaces, for unsupervised detection of dementia using cortical surface features. We utilized a pretext task of masked image modeling to learn representations of cortical surface features in a self-supervised manner and employed a novel iterative masked anomaly detection algorithm to discover deviations from those learned representations. We validate our approach in various datasets with individuals diagnosed with dementia \cite{petersen2010alzheimer,lamontagne2019oasis}. Our experiments reveal that our framework can be used to detect cortical anomalies with the following primary contributions.
\begin{enumerate}
    \item We introduced a method that learns high-dimensional abstractions from cortical features for the unsupervised detection of brain anomalies.
    \item We demonstrated the effectiveness of our framework in distinguishing dementia from healthy controls.
    \item We highlight the capabilities of our framework to identify key cortical regions involved in dementia, providing additional validation and confidence in our approach.
\end{enumerate}

%% file: 2-related.tex
\section{Related work}\label{sec:related}
\subsection{Unsupervised anomaly detection}\label{subsec:unsupervised}
Anomaly detection, a.k.a. outlier detection or novelty detection, is referred to as the process of detecting data instances that significantly deviate from the reference. Unsupervised anomaly detection (UAD) specifically aims to detect anomalies by learning the distribution of the reference samples without any labeled anomalies. This task is challenging. Classical approaches focus on using machine learning methods to build a one-class classifier, such as the one-class support vector machine \cite{scholkopf2001estimating}  \cite{tax2004support}. Other methods include statistical models that estimate probability distributions over data points, common methods include Gaussian mixture models \cite{yang2009outlier}, linear regression \cite{satman2013new}, Kernel Density Estimation \cite{pavlidou2014kernel}, and histogram-based statistical models \cite{goldstein2012histogram}.

With the advancement of deep learning, the most recent work develops neural networks for UAD tasks. Here, reconstruction-based methods are one of the most widely used approaches. These methods hypothesize that the networks trained in the healthy reference distribution can only reconstruct an anomaly-free distribution, leading to incomplete reconstruction for anomalous data that were not seen during training. Several common methods are proposed, such as (variational) Auto Encoders \cite{an2015variational,zhou2017anomaly,chen2018autoencoder,gong2019memorizing}, generative adversarial networks \cite{sabuhi2021applications,xia2022gan}, and diffusion models \cite{wolleb2022diffusion,pinaya2022fast,zhang2023unsupervised}. In addition to pixel reconstruction methods, some methods also focus on feature reconstruction, where features are extracted from pre-trained networks and then reconstructed \cite{rippel2021modeling,wan2022unsupervised,heckler2023exploring}. By using a powerful pre-trained network as feature extractor, these methods alleviates the need for training networks from scratch with limited possible supervision. Other methods such as memory matching utilize a memory component to store representations of normal data during training. This stored information is then used to identify anomalies in unseen data \cite{park2020learning,huyan2022unsupervised,zhang2022adaptive}. However, while they are effective for complex data such as images and time series by explicitly storing normal patterns, the size and complexity of the memory bank can impact performance and computational cost.

In contrast to the existing UAD methods developed for natural images, the field of medical imaging presents unique challenges, in terms of data variability and uncertainty of labels \cite{satizabal2019genetic,wolfers2021replicating}. Instead of aiming for general and transferable models across different imaging domains, personalized approaches are required for medical imaging. In this study, we propose a simple self-supervised objective function to enable personalized unsupervised anomaly detection specifically for cortical surfaces.

\input{flowchart.tex}

\subsection{Anomaly Detection in Brain Images}\label{subsec:brain}
UAD in brain images is a critical task in medical imaging, as it can help identify abnormalities in the brain that may be indicative of a disease or disorder. Several different modalities of brain imaging are used in medical imaging, including structural magnetic resonance imaging (MRI), computed tomography (CT), and positron emission tomography (PET). Each of these modalities has its own strengths and weaknesses, and different modalities may be used depending on the specific clinical question being addressed. For example, structural MRI is often used to visualize the soft tissues of the brain. Bercea et al \cite{bercea2022federated}, proposed an unsupervised deep convolutional autoencoder for multiple sclerosis, vascular lesions, and low and high grade tumors / glioblastoma. Similarly, Luo et al \cite{luo2023unsupervised} presented a three-dimensional deep autoencoder network using T2w volumes for the detection of glioblastoma, multiple sclerosis, and cerebral infarction. CT is better suited for visualizing bone structures and is often used in cases of head trauma or suspected skull fracture \cite{sato2018primitive,pinaya2022fast}. PET imaging can be used to visualize metabolic activity in the brain, which can be useful for detecting areas of increased glucose metabolism in the brain, which may be indicative of a tumor \cite{baydargil2021anomaly,solal2024leveraging}. Choi et al. \cite{choi2019deep} explore the use of VAE models for Alzheimer's detection in brain PET images. However, the task of anomaly detection in brain images is challenging due to the complexity and variability of brain anatomy, and learning a model from scratch with raw volumetric data can be computationally expensive and usually results in unsatisfactory performance. Therefore, to bridge the gap between traditional neuroscience knowledge and anomaly detection, we propose a self-supervised anomaly detection method using brain cortical surface data, which has been shown to be a useful descriptor in conventional brain imaging analysis \cite{querbes2009early,schwarz2016large,hagler2019image}.

%% file: flowchart.tex
\begin{figure*}[ht!]
    \centering
    \includegraphics[width=\textwidth]{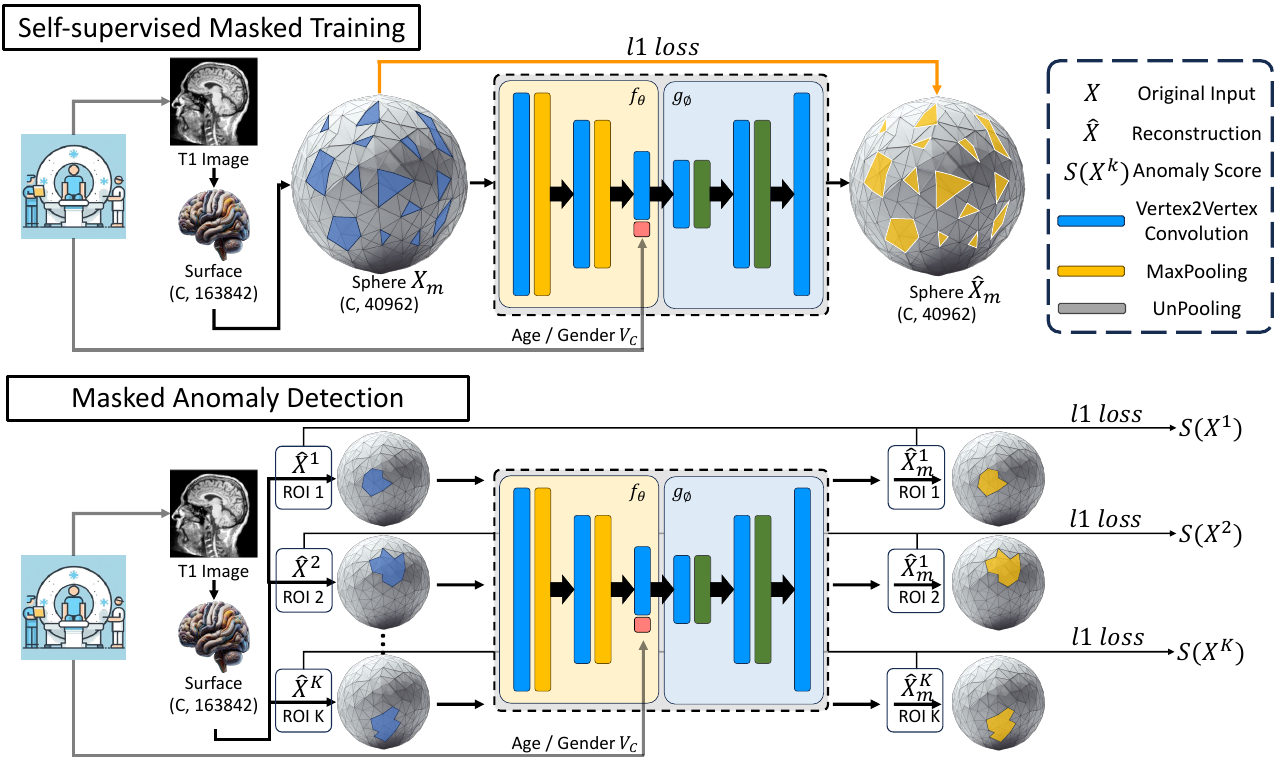}
    \caption{Our proposed Masked Mesh Net (MMN) framework.}
    \label{fig:flowchart}
\end{figure*}

%% file: 3-methods.tex
\section{Methods}\label{sec:methods}
In this section we will detail how we used self-supervised learning to train a deep model for unsupervised anomaly detection across cortical surfaces. We will first describe the data preprocessing and model architecture, followed by self-supervised masked training and unsupervised masked anomaly detection.

\subsection{Data preprocessing}
\label{sec:preprocessing}
We first extract surface features from T1 MRI images using the Freesurfer \textit{recon-all} pipeline (version 6.0) with default parameters. This process will generate $4$ mesh features: \textit{Curvature}, \textit{Sulcus}, \textit{Thickness}, and \textit{Volume}, with corresponding Desikan-Killiany atlas \cite{fischl2012freesurfer}. Subsequently, we re-tessellate all individual meshes using barycentric interpolation, from their template resolution ($163842$ vertices) to a sixth-order icosphere ($40962$ equally spaced vertices). We will also re-tessellate the Desikan-Killiany atlas to the same resolution for ROI-based inference during testing and evaluation. This step reduces the size of the mesh to a more manageable size while preserving the original surface features. Until now, we have preprocessed the data into 3D cortical surface meshes, which we will use for training and evaluation. Note that our method will not require further registration or any atlas-based division.

\subsection{Self-Supervised Masked Mesh Training}
\label{sec:masked_training}
Masked Image Modeling \cite{xie2022simmim} for self-supervised representation learning (SSL) reconstructs the randomly masking out portions of the input image and then trains the model to reconstruct the original image from the visible regions. By solving this pretext task of reconstructing the missing regions, the model learns meaningful visual representations and has been shown to be effective in various downstream tasks \cite{hsu2021hubert,chen2023masked,dong2024simmtm}. We adapted this idea to the mesh data domain and proposed a self-supervised masked mesh training method for unsupervised anomaly detection on brain cortical surfaces. The model was trained to predict the masked patches based on the context of other visible patches, which encouraged the model to learn the spatial relationships among different regions of the brain.

Given input mesh data $\mathbf{X} \in \mathbb{R}^{C \times 40962}$, where $\mathrm{C}$ is the number of channels (selected from extracted Freesurfer output: Curvature, Sulcus, Thickness, Volume) we first randomly mask out the $M$ vertices and replace them with learned mask tokens to obtain $\mathbf{X}_m$. An encoder model $f_\theta$ is then utilized to extract features:
\begin{equation}
    \mathbf{Z}^{l+1} = f_\theta^l(\mathbf{Z}^l), \quad \mathbf{Z}^0 = \mathbf{X}_m
\end{equation}
where $l$ denotes the layer index and $\mathbf{Z}^l$ is the output feature of the $l$-th layer. The output features $\mathbf{Z}^L \in \mathbb{R}^{C_{L} \times D}$, where $D$ is the hidden feature dimension, will be used to predict the masked patches. A context embedding vector $\mathbf{V_{C}}$ of phenotypic information of the subject padded from dimension $1 \times V_{C}$ to $C_{L} \times V_{C}$, where $C_{L}$ is the number of channels in the bottleneck layer of the mesh, is additionally concatenated with the bottleneck feature $\mathbf{Z}^L$:
\begin{equation}
    \mathbf{Z}^{LC} = \mathbf{Z}^L \oplus \mathbf{V_{C}}
\end{equation}
where $\mathbf{V_{C}}$ are the phenotypic features of the subject such as age and gender. The masked vertices are then predicted through a decoder model $g_{\phi}$:
\begin{equation}
    \hat{\mathbf{X}}_m = g_{\phi}(\mathbf{Z}^{LC}), \forall m \in M
\end{equation}
where $m$ denotes the embedding to the $m$-th masked vertices. Finally, the parameters $\theta$ and $\phi$ were trained by minimizing the $\ell_1$ loss between $M$ predicted and ground-truth masked vertices:
\begin{equation}
    \mathcal{L}(\theta, \phi) = \frac{1}{|M|} \sum_{m \in M} | \hat{\mathbf{X}}_m - \mathbf{X}_m |_1
\end{equation}
To this point, we have provided the model with the masked input data and the subject-level information, and the model is trained to predict the masked vertices based on the context of other visible vertices and subject-level phenotypic records. This framework compelled the model to make predictions based on spherical-spatial contextual information, encouraging the learning of complex natural patterns among brain cortical features. We will now elaborate on how we utilized the trained model for unsupervised anomaly detection.

\input{masked_detection.tex}
\subsection{Unsupervised Masked Anomaly Detection}\label{subsec:unsupervised_anomaly_detection}
After pre-training the encoder $f_\theta$ and decoder $g_\phi$ using the self-supervised masked mesh training method described in Section \ref{sec:masked_training}, we apply it for unsupervised anomaly detection on new data samples $\mathbf{X}$. We propose a masked anomaly detection approach, as outlined in Algorithm \ref{algo:masked_anomaly_detection}. The core idea of this algorithm is to mask the region of interest (ROI) in the input data $\mathbf{X}$ and reconstruct the masked vertices based on the context of other visible vertices. The final anomaly score $S(\mathbf{X})$ is calculated as the distance $\ell_1$ between the original data and the corresponding reconstructed data. Since the model was trained to predict the masked vertices based on the context of other visible vertices, a higher anomaly score will indicate a higher likelihood of the sample being anomalous compared to the training in healthy individuals. Moreover, instead of computing the anomaly score for the entire brain as a whole, we compute the anomaly score for each ROI in the Desikan-Killiany atlas. During data preprocessing as described in Section \ref{sec:preprocessing}, we re-tessellate the Desikan-Killiany atlas to the same resolution as the cortical surface meshes. Hence we are able to mask preciese ROIs in vertices space and compute anomaly scores for each ROI. This equivalently calculates a conditional anomaly score given subject's unmmasked brain regions with phenotypic information for each ROI, which automatically enable a \textit{Subject-Level Adaptation} during model inference without the need of retraining the model. This allows us to identify specific brain regions that are anomalous in the subject, which will improve the reliability and interpretability of the model.

\subsection{Model Architecture}\label{subsec:model-architecture}
Our model architecture is inspired by the work of Lei et al.\cite{lei2023mesh}. We adopt their vertex2vertex convolution and the pooling / unpooling block as the fundamental building unit in our U-Net\cite{ronneberger2015u} like the encoder-decoder architecture for MMN.

\subsubsection{Vertex2Vertex Convolution}\label{subsubsec:vertex2vertex-convolution}
The vertex2vertex convolution is a composite operation designed to effectively propagate information throughout the mesh structure. It consists of two sequential steps: a vertex2facet convolution followed by a facet2vertex convolution. This two-step process allows for comprehensive feature learning that accounts for both local vertex properties and the geometric relationships between vertices and their adjacent facets.

\paragraph{Vertex2Facet Convolution}
The vertex2facet convolution aggregates features from the vertices of a facet to compute a new feature at the facet level. Given a facet with vertices $v_1, v_2, v_3$ and their corresponding features $h_1, h_2, h_3$, the facet feature $g_f$ is computed as:
\begin{equation}
    g_f = F(\frac{\pi}{2}, 0)h_1 + F(\frac{\pi}{2}, \frac{\pi}{2})h_2 + F(0, 0)h_3
\end{equation}
where $F(\theta, \phi)$ is a filter function modeled using spherical harmonics. The angular arguments $(\theta, \phi)$ are derived from the projected Barycentric coordinates of the vertices.

\paragraph{Facet2Vertex Convolution}
Following the vertex2facet operation, the facet2vertex convolution propagates information from the facets back to the vertices. For a vertex $v$ with adjacent facets $\mathcal{N}(v)$, the updated vertex feature $g_v$ is computed as:
\begin{equation}
    g_v = \frac{1}{|\mathcal{N}(v)|} \sum_{f \in \mathcal{N}(v)} F(\theta_f, \phi_f) \cdot h_f
\end{equation}
Here, $h_f$ represents the features of facet $f$ computed in the vertex2facet step. The angles $\theta_f$ and $\phi_f$ are derived from the normal facet.

\paragraph{Filter Function}
In both operations, the filter function $F(\theta, \phi)$ is defined using truncated spherical harmonics:
\begin{equation}
    \begin{split}
        F(\theta, \phi) =
        &\sum_{l=0}^L \Big(\sum_{m=1}^l a_{lm} Y_l^m(\theta, 0)\cos(m\phi) \\
        &+ \sum_{m=1}^l b_{lm} Y_l^m(\theta, 0)\sin(m\phi) \\
        &+ a_{l0} Y_l^0(\theta, \phi)\Big)
    \end{split}
\end{equation}
where $Y_l^m$ are spherical harmonics, $L$ is the maximum degree of the spherical harmonics used, and $a_{lm}, b_{lm}$ are learnable parameters. The vertex2vertex convolution, by combining these two operations, allows for effective feature propagation across the mesh structure while maintaining geometric awareness. This makes it particularly suitable for learning hierarchical representations of 3D mesh data in our U-Net like architecture.

\subsubsection{Mesh Pooling and Unpooling}\label{subsubsec:pooling-unpooling}
For mesh pooling, we employ the GPU-accelerated mesh decimation technique in \cite{lei2023mesh}. The decimation process reduces the number of vertices while preserving the mesh structure. The pooling operation then aggregates features from the original vertices to the decimated vertices.

Let $\text{VCluster}$ be the information about the clustering of the vertex from decimation. The max pooling operation for a cluster $C$ is defined as:
\begin{equation}
    h_C = \max_{v \in C} h_v
\end{equation}
where $h_v$ are the features of vertex $v$ in the original mesh.

For unpooling, we use the reverse mapping to propagate features from the decimated mesh back to the original resolution:
\begin{equation}
    h_v = h_C, \quad \forall v \in C
\end{equation}
where $h_C$ is the feature of the cluster $C$ in the decimated mesh.

\subsubsection{Network Architecture}\label{subsubsec:network-architecture}
Our encoder consists of 2 vertex2vertex convolution blocks, each followed by a mesh pooling operation:
\begin{equation}
    H_{l+1} = \text{Pool}(\text{ConV2V}_{l2}(\text{ConV2V}_{l1}(H_l)))
\end{equation}
where $H_l$ are the features at layer $l$, $\text{ConV2V}$ is the vertex2vertex convolution, and $\text{Pool}$ is the mesh pooling operation.

The decoder mirrors this structure, using vertex2vertex convolution blocks followed by mesh unpooling operations:
\begin{equation}
    H_{l-1} = \text{ConV2V}_{l2}(\text{ConV2V}_{l1}(\text{Unpool}(H_l)))
\end{equation}
By combining the strengths of the vertex2vertex convolution with the proven effectiveness of the U-Net architecture, our model is well-suited for learning complex spatial relationships in cortical surfaces.

%% file: masked_detection.tex
\begin{algorithm}
    \caption{Masked Anomaly Detection}\label{algo:masked_anomaly_detection}
    \begin{algorithmic}[1]
        \Require \\
        Data samples $\mathbf{X}$, phenotype embedding vector $V_{C}$\\
        encoder $f_\theta$, decoder $g_\phi$,\\
        number of ROIs=K
        \Ensure Anomaly scores $S(\mathbf{X})$.
        \For{$k=1$ to $K$}
        \State Mask ROI k's vertices in $\mathbf{X}$ to obtain $\mathbf{X}^k_m$
        \State Obtain reconstructions: $\hat{\mathbf{x}}^k_m \gets g_{\phi}(f_\theta(\mathbf{X}^k_m), V_{C})$
        \State Compute the anomaly score: $S(\mathbf{X}^{k}) = ||\mathbf{x}^k_m - \hat{\mathbf{x}}^k_m||_1$.

        \EndFor
    \end{algorithmic}
\end{algorithm}

%% file: 4-experiments.tex
\section{Experiments}\label{sec:experiments}
\subsection{Experimental Setup}\label{subsection:experimental-setup}
\subsubsection{Dataset}\label{subsubsection:dataset}
Table \ref{table:demographics} summarizes the demographics of all datasets used in the experiments. For self-supervised model training, we compose a dataset of $20083$ subjects from the UKB\cite{miller2016multimodal} and HCP\_Aging\cite{bookheimer2019lifespan} datasets. Specifically, we excluded subjects in UKB dataset by the exclusion criteria: 1) subjects with missing age or gender 2) subjects with neurologic or psychiatric conditions in ICD-10 codes: ['A8', 'B20', 'B21', 'B22', 'B23', 'B24', 'B65-B83', 'C', 'F', 'G', 'I6', 'Q0','S04', 'S06', 'S07', 'S08', 'S09', 'T36-T46', 'T48-T50'] 3) subjects with obvious neurological imaging defects in ICD-10 codes: ['C70', 'C71', 'C72', 'F2', 'F31', 'F7', 'G', 'I6', 'Q0', 'R90', 'R940', 'S01', 'S02', 'S03', 'S04', 'S05', 'S06', 'S07', 'S08', 'S09'] This healthy control dataset is further divided into 60/40 ratio with equally distributed gender and age as training and validation set respectively. To verify our proposed unsupervised anomaly detection framework, we further organize a test set from two different sources for verification. The first one is the ADNI\cite{petersen2010alzheimer} dataset, which contains $198$ subjects with $51$ AD patients. The ADNI was launched in 2003 as a public-private partnership, led by Principal Investigator Michael W. Weiner, MD. The primary goal of ADNI has been to test whether serial magnetic resonance imaging (MRI), positron emission tomography (PET), other biological markers, and clinical and neuropsychological assessment can be combined to measure the progression of mild cognitive impairment (MCI) and early Alzheimer's disease (AD). The second one is the OASIS3\cite{lamontagne2019oasis} dataset, which contains $925$ subjects with $175$ AD patients. By combing these two datasets, we eventually have a test set of $1123$ subjects with $226$ subjects diagnosised as AD. Our model will first trained and validated on the training and validation set, and then evaluated on the labeled test set to verify the anomaly detection capability. All subjects with a median-centered absolute Euler number greater than 25 during preprocesing were excluded, as these were found to be of poor quality \cite{kia2022closing}. Fig. \ref{fig:data_pipeline} illustrates the overall data pipeline used in the experiments.
\input{demographic.tex}
\input{data_pipeline.tex}
\input{result-1way_anova_adni_oasis3.tex}

\subsubsection{Training and Evaluation Details}\label{sssec:training_and_evaluation}
In our experiments all methods were trained on the training set and underwent hyperparameter searches using the validation set. The best model was then evaluated on the held-out test set to prevent information leakage. Several hyperparameters were searched according to the original paper: number of masked patches in training $M$ = 50\% of total vertices, degree of spherical harmonics $L$ = 3, encoder channels searched among ([16, 32, 64, 128], [32, 64, 96, 128], [32, 64, 128, 256]), convolution stride set as [2, 2, 2, 2], and learning rate searched among ([1e-3, 1e-4]). All models were trained using the AdamW optimizer with a cosine annealing scheduler and early stopping with a maximum of 50 epochs. The loss function $\ell_1$ was utilized as an objective function and evaluation metric for model selection. The model was trained on two NVIDIA A100 GPU with 80GB of memory. The training time was approximately 6 hours for each model. The code implemented in PyTorch will be made available upon publication.


\subsection{Unsupervised Anomaly Detection Results}\label{subsection:unsupervised-anomaly-detection-results}
To validate the effectiveness of our proposed self-supervised masked mesh learning framework for unsupervised anomaly detection, we calculated anomaly scores for all ROIs of each subject in both ADNI and OASIS3 datasets. These scores were then used to classify subjects into normal and abnormal (Alzheimer's Disease) groups. We used one-way ANOVA to evaluate the differences between these groups, followed by the Benjamini-Hochberg procedure\footnote{The Benjamini-Hochberg procedure was implemented using statsmodels \cite{seabold2010statsmodels}.} for the correction of multiple p-value tests. We plotted the Effect Size (Eta Squared, $\eta^2$) for all regions with corrected p-values\footnote{Detailed p-values are available in the supporting documents/multimedia tab.} $< 0.05$. Figures \ref{fig:result_1way_anova_eta_adni_oasis3} summarize these results.
Several key findings emerged:
\begin{enumerate}
    \item Among the four cortical features examined, cortical thickness (\textit{Thickness} in both figures) proved to be the most effective for anomaly detection. This aligns with the existing literature that identified cortical thickness as a sensitive biomarker of AD \cite{querbes2009early,schwarz2016large,holbrook2020anterolateral,longhurst2023cortical}. Furthermore, the left hemisphere generally demonstrated higher sensitivity in detecting anomalies compared to the right hemisphere.
    \item Several cortical regions consistently showed effectiveness in anomaly detection in both datasets: \begin{itemize}
              \item Left hemisphere cortical thickness: superior frontal, precentral, and  transverse temporal regions
              \item Right hemisphere cortical thickness: precentral regions
              \item Left hemisphere cortical sulcus: fusiform, postcentral and parahippocampal regions
              \item Right hemisphere cortical sulcus: supramarginal region
              \item Left hemisphere cortical volume: lateral orbitofrontal, rostral middle frontal, and pars triangularis regions
              \item Right hemisphere cortical volume: lateral orbitofrontal and pars triangularis regions
          \end{itemize}
          These regions have previously been associated with AD in various studies \cite{ikonomovic2007superior,hallam2020neural,du2023association,salat2009regional,devanand2012mri,leandrou2020assessment,yeung2021anterolateral}.
\end{enumerate}
In conclusion, these results demonstrate the efficacy of our proposed self-supervised masked mesh learning framework in detecting anomalies associated with AD patients.

%% file: demographic.tex
\begin{table}[t!]
    \caption{Demographics of the datasets used in the study.}
    \resizebox{\columnwidth}{!}{%
        \begin{tabular}{@{}ccccc@{}}
            \toprule
            \multicolumn{5}{c}{\normalsize{\textbf{Training Set}}}                                              \\ \midrule
            Dataset                                 & \# Subjects & \# Patients & Age (mean±std) & Gender (M/F) \\ \midrule
            UKB\cite{miller2016multimodal}          & 11709       & -           & 63.33 ± 7.43   & 5427 / 6282  \\
            HCP\_Aging\cite{bookheimer2019lifespan} & 342         & -           & 56.79 ± 13.74  & 148 / 194    \\ \midrule
            \multicolumn{5}{c}{\normalsize{\textbf{Validation Set}}}                                            \\ \midrule
            UKB\cite{miller2016multimodal}          & 7806        & -           & 63.44 ± 7.47   & 3598 / 4208  \\
            HCP\_Aging\cite{bookheimer2019lifespan} & 226         & -           & 58.04 ± 14.66  & 93 / 133     \\ \midrule
            \multicolumn{5}{c}{\normalsize{\textbf{Test Set}}}                                                  \\ \midrule
            ADNI\cite{petersen2010alzheimer}        & 198         & AD (51)     & 75.49 ± 7.13   & 96 / 102     \\
            OASIS3\cite{lamontagne2019oasis}        & 925         & AD (175)    & 69.54 ± 9.55   & 527 / 398    \\ \bottomrule
        \end{tabular}
    }
    \label{table:demographics}
\end{table}

%% file: data_pipeline.tex
\begin{figure}[t!]
    \centering
    \includegraphics[width=\columnwidth]{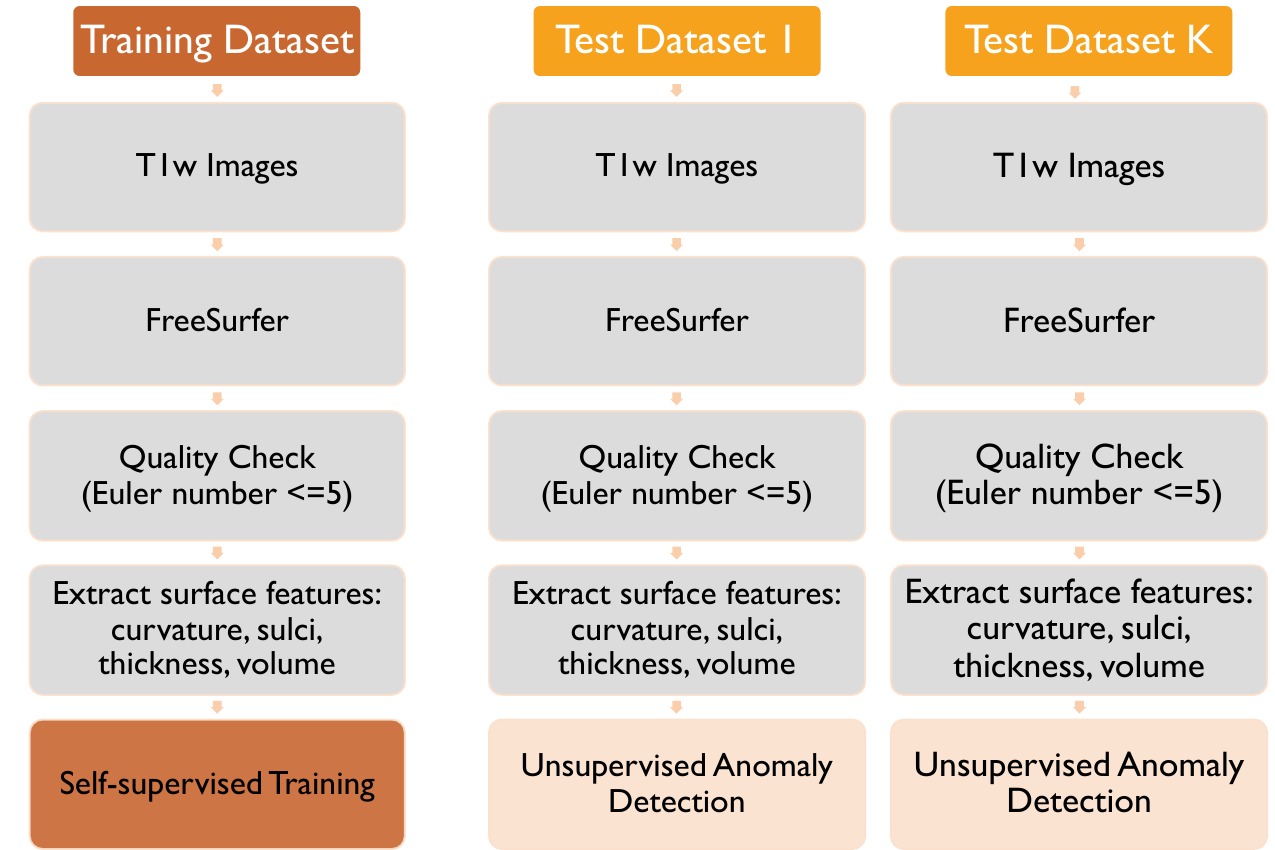}
    \caption{Data pipeline for unsupervised anomaly detection.}
    \label{fig:data_pipeline}
\end{figure}

%% file: result-1way_anova_adni_oasis3.tex
\begin{figure*}[ht!]
    \centering
    \includegraphics[width=0.9\textwidth]{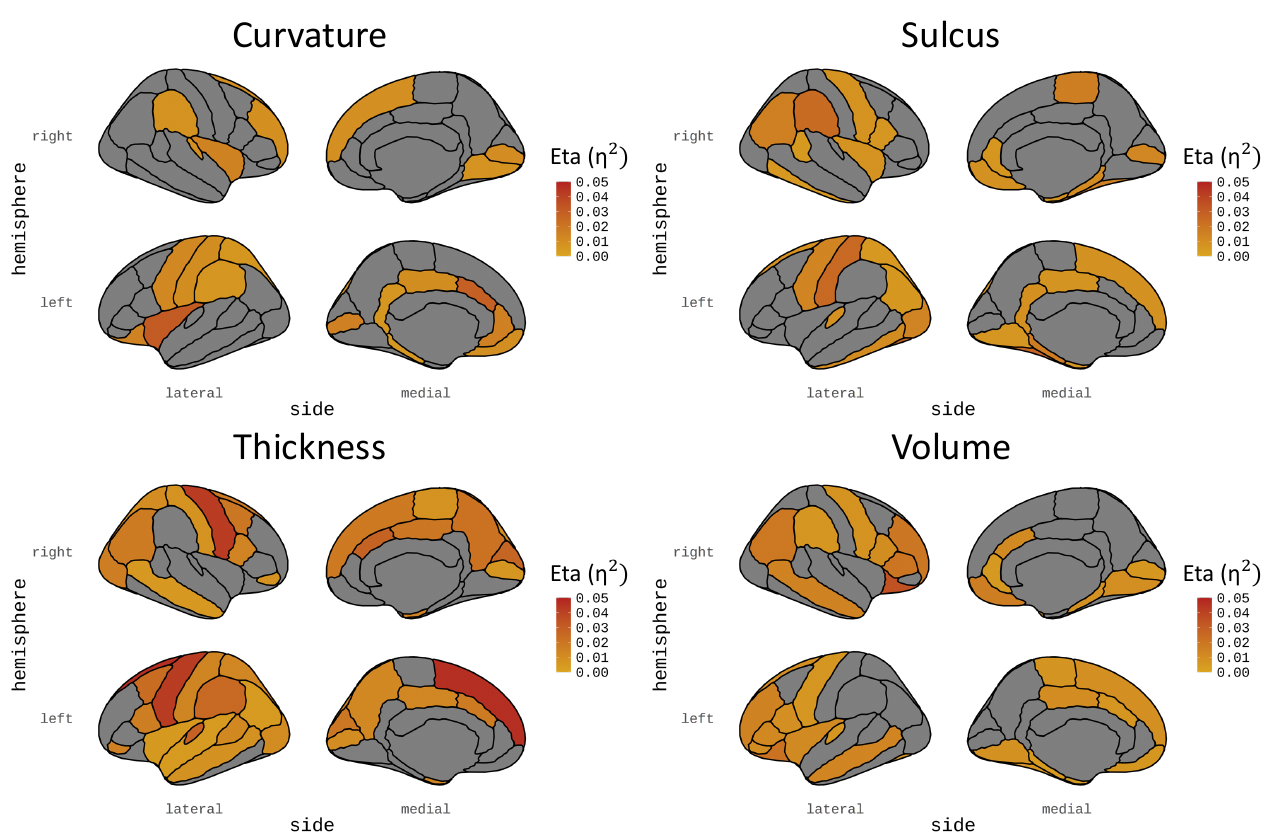}
    \caption{Unsupervised anomaly detection result in test set.}
    \label{fig:result_1way_anova_eta_adni_oasis3}
\end{figure*}

%% file: 5-discussion.tex
\section{Discussion}\label{sec:discussion}

This study introduces a novel self-supervised masked mesh learning (MMN) framework for unsupervised anomaly detection on 3D cortical surfaces. This efficient approach learns representations of normal cortical structure variability from healthy reference datasets, including the UKB \cite{miller2016multimodal} and HCP-Aging \cite{harms2018extending,bookheimer2019lifespan} datasets, by predicting masked regions of the cortical surface. The anomaly detection in unseen participants is then performed by calculating reconstruction errors from the MMN, offering a generalizable method for individual-level analysis. While previous studies have focused on supervised learning approaches for detecting cortical anomalies \cite{rallabandi2020automatic,shin2021cortical,pateria2024comprehensive}, our self-supervised MMN framework provides a more flexible and scalable approach that does not require labeled data, making it suitable for rare diseases and complex disorders characterized by cognitive decline \cite{decherchi2021opportunities,dos2022towards,chen2023algorithmic,knopman2003essentials,pike2022subjective}. Our method leverages the rich geometric features of the cortical surface to detect subtle structural changes that may signify the early stages of dementia and cognitive decline \cite{im2008sulcal,tang2021slower,blanc2015cortical,vuksanovic2019cortical,thompson2007tracking}. By focusing on the cortical surface, our method successfully identified anomalies in cortical thickness, volume, and sulcal features, aligning with known patterns in Alzheimer's disease \cite{schwarz2016large,chandra2019magnetic,holbrook2020anterolateral}. This highlights the potential of our approach to detect subtle morphological deviations related to neurological and age-related conditions. We focus on Alzheimer's disease because early detection of subtle changes in brain structure and function is crucial to understanding the onset and progression of the disease \cite{filippi2020changes,ferrari2021complexity,leonardsen2024constructing}.

Alzheimer's disease is characterized by progressive neurodegeneration, leading to changes in cortical thickness, volume, and other structural features. Effective anomaly detection can reveal early biomarkers, track disease progression, and support timely diagnosis and intervention \cite{rasmussen2019alzheimer,porsteinsson2021diagnosis,vogel2022subtypes}. Therefore, integrating anomaly detection methods holds significant promise in improving diagnostic precision and advancing personalized treatment \cite{reitz2016toward,hampel2019alzheimer,therriault2024biomarker}. We evaluated our method using data from the Alzheimer's Disease Neuroimaging Initiative (ADNI) and the Open Access Series of Imaging Studies (OASIS3), both of which provide extensive biomarker data for training and validation. Our results demonstrate robust associations with the disease, improving our understanding of cortical changes in various stages of abnormal aging.

\subsection{Limitations}\label{subsec:limitations}
Cortical anomalies, although indicative, are not the only markers of Alzheimer's Disease \cite{blennow2018biomarkers}. It also involves significant changes in subcortical structures and cellular processes that are not captured by our current method or readily detectable with current imaging technology. Although our framework is optimized for cortical anomalies, it could be adapted to incorporate other features based on mesh preprocessed data \cite{schwartz2023evolution,demirci2023systematic,kalantar2023cross}, extending its applicability. Furthermore, incorporating cognitive functions closely related to cortical regions, such as language and executive function, could provide further insight as to the meaning that cortical abnormalities hold in the context of cortical functions. This flexible framework could be extended to various data sources, offering a comprehensive approach to studying cortical and cognitive changes in Alzheimer's disease and other neurological and psychiatric conditions involving abnormal cortical function.
Our reliance on reconstruction error as the primary metric for anomaly detection may present limitations, as it might not fully capture subtle or complex anomalies on the cortical surfaces \cite{darayi2022computational,schwartz2023evolution}. Reconstruction error measures the discrepancy between the original and reconstructed data, which is useful for identifying obvious deviations from the norm. However, subtle or intricate anomalies, which might be indicative of early or nuanced pathological changes, may not significantly affect the overall reconstruction error or anomalies that are spread over space \cite{lamballais2020cortical,demirci2023systematic}. This limitation suggests that, while the reconstruction error is a valuable metric, it may not be sufficient on its own for a comprehensive detection of all types of anomalies. As our work is focused on cortical surfaces, future research should explore additional metrics and analytical approaches that can complement the reconstruction error. Metrics such as feature-based similarity measures, statistical outlier detection, or domain-specific anomaly scores could help identify subtle or complex anomalies in future \cite{tschuchnig2022anomaly}. 
Finally, our approach demands significant computational resources, potentially limiting accessibility for research groups with limited high-performance computing infrastructure and memory. However, we share our models with other scientists and after training the computational requirements are limited and can be executed on a local machine. 

\subsection{Future Directions}\label{subsec:comparison-with-different-hyperparameters}
In future work we intend to investigate the development of the cortex throughout the entire life of the brain, rather than focus solely on aging. By incorporating a lifespan reference dataset that spans different life stages, from childhood through adulthood to late age, we can develop a framework that can detect anomalies conditional on different life stages to facilitate our understanding of cortical changes and their implications for various developmental and degenerative conditions. This approach will enable us to model the development of the cortex over the lifespan and improves the accuracy of anomaly detection and increases the availability of advanced modeling techniques in this research area. In general, our self-supervised masked mesh learning framework represents an advancement in unsupervised anomaly detection for brain surfaces, opening up new research and clinical applications to understand a diverse set of cortical anomalies. Furthermore, it is essential to expand our research to include other types of neurological and psychiatric disorders, such as schizophrenia. Schizophrenia, which often appears in the late teens to early twenties, presents distinct neuroanatomical and functional abnormalities that are not fully captured by data sets primarily focused on aging \cite{kalantar2023cross,schwartz2023evolution}. This expansion would allow us to tailor and validate our self-supervised masked mesh learning framework to detect anomalies associated with other medical conditions \cite{banaj2023cortical,yang2024learning}. By adapting our approach to account for these early-onset conditions, we can enhance the model's ability to identify and differentiate between various types of brain abnormalities across a broader range of neurological and mental disorders. Understanding how these disorders manifest in different life stages and how they affect brain structures differently from conditions like Alzheimer's disease would provide a more comprehensive view of the development of these saver medical conditions.

%% file: 7-conclusion.tex
\section{Conclusion}\label{sec:conclusion}
This paper tackles the complex challenge of UAD on the cortical surfaces of the human brain by presenting an innovative self-supervised masked mesh learning framework tailored specifically for 3D cortical surfaces. Traditional methods in anomaly detection often struggle with the high dimensionality and intricate geometry of cortical surfaces. To address this, our framework leverages the unique intrinsic geometry of the cortical surface, utilizing a masked mesh convolutional neural network designed to predict masked regions and effectively capture normal variations. This approach allows the model to learn from a large dataset of healthy subjects, enabling it to distinguish between typical and atypical features based on observed patterns of cortical morphology. The framework's effectiveness was rigorously validated through its application to Alzheimer's disease, a condition known for its distinct impact on cortical structures. Our method demonstrated a robust ability to identify anomalies in various cortical features, including thickness, volume, and sulcus patterns, which are critical for understanding Alzheimer's disease. By comparing detected anomalies with known pathological changes, the framework validated its capacity to uncover subtle deviations linked to the disease. In general, our self-supervised masked mesh learning framework represents a significant advance in the field of brain imaging and anomaly detection. It introduces a novel approach to harnessing the geometric properties of cortical surfaces for the improved detection of abnormalities. The framework not only improves the accuracy of detecting cortical anomalies but also paves the way for further research into its application across different neurological conditions, potentially leading to more refined diagnostic tools and a deeper understanding of the developmental and degenerative diseases of the cortex.